\documentclass[10pt,conference]{IEEEtran}

\usepackage{graphicx,epsfig,amssymb,amsmath,url,latexsym,stfloats,array,bm,bbm}
\usepackage[tight,footnotesize]{subfigure}
\usepackage{makecell}
\usepackage{mathrsfs}
\usepackage{amsthm}
\usepackage{longtable}
\usepackage{amsfonts}
\usepackage{caption}
\usepackage{graphicx}
\usepackage{multirow}
\usepackage{longtable}
\usepackage{afterpage}
\usepackage{subeqnarray}
\usepackage{epstopdf}
\usepackage{empheq}
\usepackage{latexsym}
\usepackage{color, soul}
\usepackage{framed}
\usepackage{lipsum}
\usepackage{color}
\usepackage{stfloats}
\usepackage{setspace}
\definecolor{shadecolor}{rgb}{1,0,0}
\usepackage{cite}
\usepackage{hyperref}
\usepackage{verbatim}
\usepackage{algorithmic}
\usepackage[linesnumbered,ruled]{algorithm2e}
\usepackage[hang]{footmisc}
\usepackage{caption}
\usepackage{cuted}
\newcommand{\HRule}[1][\medskipamount]{\par
	\vspace*{\dimexpr-\parskip-\baselineskip+#1}
	\noindent\rule{\linewidth}{0.2mm}\par
	\vspace*{\dimexpr-\parskip-1\baselineskip+#1}}

\begin{document}
\title{
Personalized QoE Enhancement for Adaptive Video Streaming:  A Digital Twin-Assisted Scheme}
{\setstretch{1.0}
	\author{
	\IEEEauthorblockN{Xinyu Huang\IEEEauthorrefmark{1}, Conghao Zhou\IEEEauthorrefmark{1},  Wen Wu\IEEEauthorrefmark{2}, Mushu Li\IEEEauthorrefmark{1}, Huaqing Wu\IEEEauthorrefmark{1}, and Xuemin (Sherman) Shen\IEEEauthorrefmark{1}}
	    \IEEEauthorblockA{\IEEEauthorrefmark{1}Department~of~Electrical~\&~Computer~Engineering,~University~of~Waterloo,~Canada
	    \\\IEEEauthorrefmark{2}Frontier Research Center, Peng~Cheng~Laboratory,~China
	    \\Email: \{x357huan, c89zhou, m475li, h272wu, sshen\}@uwaterloo.ca, wuw02@pcl.ac.cn}
			}
}

\maketitle
\pagestyle{empty}  
\thispagestyle{empty} 
\begin{abstract}
In this paper, we present a digital twin (DT)-assisted adaptive video streaming scheme to enhance personalized quality-of-experience (PQoE). Since PQoE models are user-specific and time-varying, existing schemes based on universal and time-invariant PQoE models may suffer from performance degradation. To address this issue, we first propose a DT-assisted PQoE model construction method to obtain accurate user-specific PQoE models. Specifically, user DTs (UDTs) are respectively constructed for individual users, which can acquire and utilize users’ data to accurately tune PQoE model parameters in real time. Next, given the obtained PQoE models, we formulate a resource management problem to maximize the overall long-term PQoE by taking the dynamics of users’ locations, video content requests, and buffer statuses into account. To solve this problem, a deep reinforcement learning algorithm is developed to jointly determine segment version selection, and communication and computing resource allocation. Simulation results on the real-world dataset demonstrate that the proposed scheme can effectively enhance PQoE compared with benchmark schemes.
	
\end{abstract}

\section{Introduction}
With the rapid popularization of emerging video applications, video streaming data accounts for the majority of global mobile data \cite{Sun}, which has placed a growing strain on wireless networks. To guarantee the user's continuous playback under dynamic channel conditions, adaptive bitrate (ABR) technology \cite{Liu} that splits a complete video sequence into multiple segments with different bitrates enables adaptive video streaming. However, the traditional ABR schemes usually adopt users' predicted throughput as the principle of bitrate selection, which may not improve users' personalized watching experience that depends on multiple factors \cite{Yin}. To handle this issue, the personalized quality-of-experience (PQoE) model is proposed, which can characterize the user-specific perception of different QoE factors, including rebuffer time, video quality, and quality variation \cite{Wang_Y}. Based on the PQoE model, appropriate segments and network resources are allocated to the user to enhance its personalized watching experience. 

In the literature, significant research efforts have been put to construct the PQoE model. Wang \textit{et al.} proposed a PQoE construction method that combined user’s service preferences and network layer configurations, which can effectively characterize user’s resource demands \cite{Wang_Y}. Gao \textit{et al.} enhanced the PQoE model accuracy by leveraging the sensing information from the network \cite{Gao_Y}. To perform resource management based on the constructed PQoE model, a deep reinforcement learning (DRL)-based scheme was proposed to solve the PQoE-oriented resource allocation problem \cite{Wang_F2}. The above works focus on time-invariant PQoE models. However, PQoE models may vary across different video contents and playback statuses, such as buffer occupancy, video quality, and quality variation. Such dynamics make the previous constructed accurate PQoE models outdated over time, thereby rendering sub-optimal resource management decisions and degrading PQoE performance. Hence, constructing a real-time and accurate PQoE model is paramount.

To tackle this challenge, digital twin (DT) technology is a potential solution. DT is a digital representation of a physical entity (PE) that can accurately reflect its status and feature via real-time synchronization between the DT and PE \cite{Shen_X}. DT technology has been widely applied to fault diagnosis and predictive maintenance \cite{Tom}. In adaptive video streaming, DT technology can be leveraged to store and analyze users' data, such that the accurate user-specific PQoE model can be established in real time and then utilized to make resource management decisions for enhancing PQoE.

In this paper, we present a DT-assisted adaptive video streaming scheme to enhance PQoE. Specifically, we propose a DT-assisted PQoE model construction method to tune PQoE model parameters in real time, and design a tailored DRL-based algorithm for resource management. Firstly, user DTs (UDTs) are established to construct real-time PQoE models by storing and analyzing users’ data related to adaptive video streaming. The constructed PQoE model not only adopts the linear weighting combination of QoE factors, but also incorporates the Ebbinghaus memory effect to reflect the fading effect of user's watching experience. The personalization is reflected in PQoE model parameters, i.e., relative memory length and sensitivity degree of QoE factors. The PQoE model parameters are tuned in real time by adopting a data fitting method. Secondly, we develop a DRL-based resource management algorithm to efficiently facilitate the PQoE-oriented adaptive video streaming. The objective is to maximize the overall long-term PQoE considering the dynamics of users’ locations, video content requests, and buffer statuses, by jointly optimizing the segment version selection, and communication and computing resource allocation. To reduce the algorithm training complexity, we narrow down the action dimension by splitting the range of transmission and transcoding variables into multiple parts to represent the decisions of different segment versions. Extensive simulation results on the real-world dataset demonstrate that the proposed DT-assisted scheme, including the PQoE model construction and the corresponding resource management algorithm, can effectively enhance PQoE compared with benchmarks. The main contributions of this paper are summarized as follows:

\begin{itemize}
	\item[$\bullet$] 
	We propose a DT-assisted PQoE model construction method, which can obtain user’s real-time and accurate PQoE model parameters.
	
	\item[$\bullet$] We develop a DRL-based resource management algorithm to jointly determine the segment version selection, and communication and computing resource allocation. 
	
\end{itemize}

The remainder of this paper is organized as follows. The DT-assisted adaptive video streaming scheme is presented in Section \ref{system}. The DRL-based resource management algorithm is proposed in Section \ref{Solution}. Simulation results are provided in Section \ref{Result}, followed by the conclusion in Section \ref{Conclusion}.

\section{DT-Assisted Adaptive Video Streaming Scheme}\label{system}
\begin{figure}[t]
	\centering
	\includegraphics[width=8.8cm]{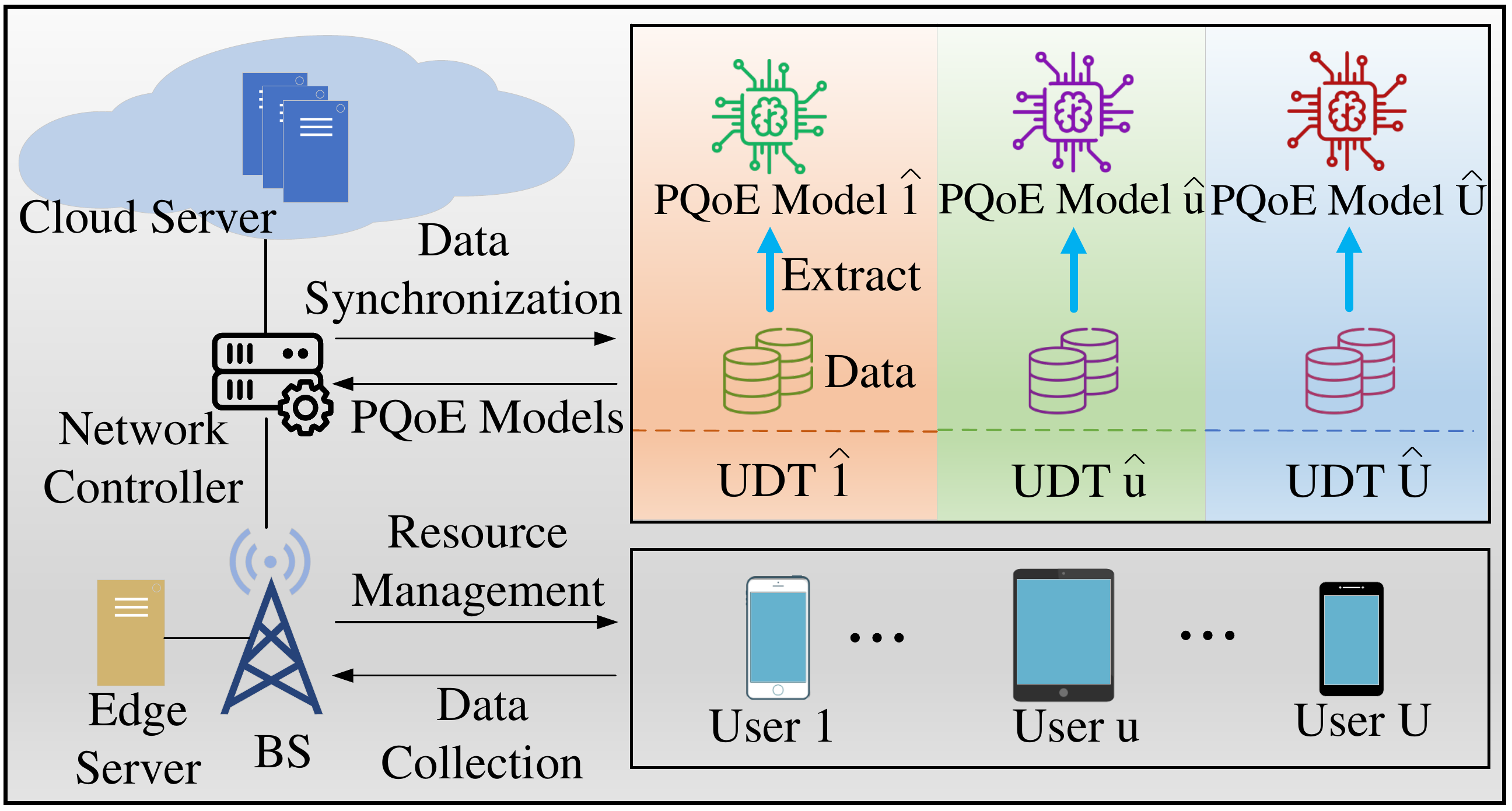}
	\caption{DT-assisted adaptive video streaming.}
	\label{fig:video_services}
\end{figure}
\subsection{System Model}
As shown in Fig.~\ref{fig:video_services}, we consider the DT-assisted adaptive video streaming framework consisting of three parts: PEs, DTs, and interaction links.

\begin{itemize}

	\item[$\bullet$] PEs: PEs consist of a base station (BS), an edge server, a cloud server, and multiple users. The set of users and the set of video sequences are denoted by $\mathcal{U}$ and $\mathcal{F}$, respectively, where $u$ and $f$ represent the corresponding indexes. Video sequences are encoded into multiple segments with different versions. The set of all segment versions is denoted by~$\mathcal{L}=\{1,...,l,...,L\}$. The edge server caches popular videos for low-latency content delivery due to limited storage capacity, while the cloud server caches all videos.
	 
	\item[$\bullet$] DTs: There are multiple UDTs deployed at the edge server. Each UDT corresponds to a user, which stores the corresponding user’s data and extracted PQoE model. The set of UDTs is denoted by $\hat{\mathcal{U}}$.
	\item[$\bullet$] Interaction links: Interaction between PEs and DTs is bidirectional coordinated by the network controller, including data synchronization and PQoE models transmission.
\end{itemize}

	The proposed framework operates as follows. When users request adaptive video streaming services, the network controller deployed at the BS will coordinate the BS, the edge server, the cloud server, and UDTs to transmit and transcode segments. Firstly, the network controller is used to collect users’ data, including requested videos, rebuffer time, video quality, quality variation, and engagement time. Specifically, the rebuffer time and engagement time data are collected from users, while the rest data are collected from the BS. Secondly, user's data are uploaded to respective UDTs through the network controller to update the stored data. Based on which, the real-time PQoE model parameters is obtained by data analysis method, which are used to update respective PQoE models. Thirdly, the real-time and accurate PQoE models in UDTs are provided to the network controller to make resource management decisions, including segment version selection, communication and computing resource allocation. Specifically, the cloud server transmits the required video segments to the BS based on the segment version selection decision. The edge server transcodes cached segments of high versions to those of targeted versions based on the computing resource allocation decision. The BS transmits the processed segments to users based on the communication resource allocation decision.

\subsection{Service Delay Model}
The adaptive video streaming system operates in each scheduling slot, indexed by $t$. The segment can be delivered to the user in the following three cases, and the corresponding service delay is analyzed.

\textbf{Case 1:} If the segment is downloaded from the edge server without transcoding, the service delay, $D_{t,u}^{(1)}$, refers to the transmission delay from the BS to user $u$ at scheduling slot $t$, given by 
\begin{equation}
		D_{t,u}^{(1)}=\sum\limits_{l=1}^{L}\frac{{g_{k_{t,u}}^{l}}\chi _{k_{t,u}}^{l}\varsigma_{k_{t,u}}^{l}}{r_{t,u}^{BS}}.
\end{equation}
Here, $k_{t,u}$ represents the next segment index of the maximum segment index at user $u$'s buffer at scheduling slot~$t$. Here, decision variable $g_{k_{t,u}}^{l}\in\{0,1\}$ indicates whether to transmit segment $k_{t,u}$ of version~$l$ to user~$u$ at scheduling slot~$t$. If the segment of version~$l$ is transmitted, $g_{k_{t,u}}^{l}=1$; Otherwise, $g_{k_{t,u}}^{l}=0$. The parameter $\chi _{k_{t,u}}^{l}\in\{0,1\}$ indicates whether segment $k_{t,u}$ of version~$l$ for user $u$ has been stored in the edge server. If segment $k_{t,u}$ of version~$l$ has been stored, $\chi _{k_{t,u}}^{l}=1$; Otherwise, $\chi _{k_{t,u}}^{l}=0$. Here, $\varsigma _{k_{t,u}}^{l}$ denotes the file size of segment $k_{t,u}$ of version $l$, and ${{r}_{t,u}^{BS}}$ denotes the average transmission capacity of user $u$ at scheduling slot~$t$. Here, ${{r}_{t,u}^{BS}}={{\xi }_{t,u}} W\log (1+{{P}_{t,u}}/{{N}_{o}})$, where ${{P}_{t,u}}$ is user $u$'s received signal power, ${{N}_{o}}$ is the noise power, and decision variable ${{\xi }_{t,u}\in\left[0,1\right]}$ represents the ratio of total bandwidth $W$ allocated to user $u$ at scheduling slot $t$.

\textbf{Case 2:} If the segment is downloaded from the edge server with transcoding, the service delay, $D_{t,u}^{(2)}$, includes transcoding delay in the edge server and the transmission delay from the BS to user $u$ at scheduling slot $t$, given by
	\begin{equation}
		\begin{aligned}
			D_{t,u}^{(2)}&=\sum\limits_{l=1}^{L}{g_{k_{t,u}}^{l}(1\!-\!\chi _{k_{t,u}}^{l})o_{k_{t,u}}^{l}}\left( \frac{\mu \varsigma_{k_{t,u}}^{l}}{{{\omega }_{t,u}}{{c}_{t}}} \!+\!\frac{\varsigma_{k_{t,u}}^{l}}{r_{t,u}^{BS}}\right).
		\end{aligned}
\end{equation}Here, decision variable $o_{k_{t,u}}^{l}\in\{0,1\}$ indicates whether segment $k_{t,u}$ of version $l$ can be obtained by transcoding. If segment $k_{t,u}$ of version $l$ can be obtained by transcoding, $o_{k_{t,u}}^{l}=1$; Otherwise, $o_{k_{t,u}}^{l}=0$.  The computing intensity for transcoding per unit file size is denoted by $\mu$, and $c_t$ is the computing capacity of the edge server at scheduling slot~$t$. Here, decision variable $\omega_{t,u}\in \left[0,1\right]$ represents the ratio of computing resources allocated to user $u$ at scheduling slot~$t$. 

\textbf{Case 3:} If the segment is downloaded from the cloud server, the service delay, $D_{t,u}^{(3)}$, refers to the transmission delay from the cloud server to the BS, and the BS to user $u$ at scheduling slot $t$, given by
	\begin{equation}
			D_{t,u}^{(3)}=\sum\limits_{l=1}^{L}{g_{k_{t,u}}^{l}(1-\chi _{k_{t,u}}^{l})
				} (1-o_{k_{t,u}}^{l})\left(\frac{\varsigma_{k_{t,u}}^{l}}{r^C} + \frac{\varsigma_{k_{t,u}}^{l}}{r_{t,u}^{BS}}\right), 
\end{equation}
where ${{r}^{C}}$ is the transmission capacity between the cloud server and the BS. Due to the wired link connection between the cloud server and the BS, ${{r}^{C}}$ is assumed to be a constant.

Based on these, service delay $D_{t,u}$ can be represented by
\begin{equation}\label{service_delay}
	D_{t,u}=D_{t,u}^{(1)} + D_{t,u}^{(2)} + D_{t,u}^{(3)},
\end{equation}
which impacts the rebuffer time in the PQoE model.

\subsection{PQoE Model}
The PQoE model consists of the following three factors.

\textbf{Rebuffer time:} The rebuffer time is related to service delay (as analyzed in Eq.~\eqref{service_delay}) and playback buffer occupancy. The playback buffer occupancy\footnote{Note that if the user stops watching the current video and switches to another video, the video packets in the current playback buffer will not be used and the playback buffer occupancy will be empty.} at the current scheduling slot depends on the playback buffer occupancy at the previous scheduling slot, current received segment time length, and scheduling slot length. Let $B_{t,u}$ denote user $u$'s playback buffer occupancy at scheduling slot $t$, which is updated via
\begin{equation}\label{B_t}
	{{B}_{t+1,u}}={{\left( {{B}_{t,u}}+ \sum\nolimits_{l=1}^{L}{g_{k_{t,u}}^{l}}e-d \right)}^{+}}.
\end{equation}
Here, $e$ is the time length of segment\footnote{The video sequence is encoded into multiple equal-length segments.}, and $d$ is the scheduling slot length. The function ${{(x)}^{+}}=\max \{x,0\}$. 

When the user's service delay exceeds the current buffer occupancy, the rebuffering event occurs. Therefore, the rebuffer time of user $u$ at scheduling slot~$t$ is defined by
\begin{equation}
	R_{t,u} = \left(D_{t,u}-B_{t,u}\right)^{+}.
\end{equation}

\textbf{Video quality:} Since one video sequence consists of multiple video segments with different video qualities, the higher video quality can usually bring the higher PQoE to user. At scheduling slot~$t$, the video quality of the new segment added to user $u$'s playback buffer is defined by
\begin{equation}
\begin{aligned}
V{_{t,u}} &=\sum\nolimits_{l=1}^{L}{g_{k_{t,u}}^{l}{\wp}_{k_{t,u}}^{l}}.
\end{aligned}
\end{equation}
Here, ${\wp}_{k_{t,u}}^{l}$ is the peak signal-to-noise ratio (PSNR) of segment $k_{t,u}$. 

\textbf{Quality variation:} Since the content-coding complexity of adjacent segments is different, segments with the same version also exhibits different PSNRs \cite{Huang_X_1}. However, in this case, the user cannot feel the quality variation. Therefore, we use switching magnitude of segment versions to represent quality variation, which is defined by
\begin{equation}
{H_{t,u}}=\sum\nolimits_{l=1}^{L}{g_{k_{t,u}}^{l}}|{{l}_{k_{t,u}}}-{{l}_{k_{t,u}-1}}|.
\end{equation}

Based on these factors, we can formulate the PQoE model. Specifically, since video playback is usually a long-term process for users, the impact of previous bad or good watching experiences, such as rebuffer time, video quality, and quality variation, on users' current watching experience decreases gradually. Therefore, we incorporate the Ebbinghaus memory effect \cite{Duanmu} into the linear weighting combination of PQoE factors \cite{Yin} to obtain an accurate PQoE model, i.e.,
\begin{equation}\label{QoE_1}
\begin{aligned}
{Z_{t,u}}({\mathbf{g}_{t,u}},{\mathbf{o}_{t,u}},{{\omega }_{t,u}},{{\xi }_{t,u}})&=\exp \left(-\frac{({p}_u-t)}{{{\lambda }_{t,u}}}\right)[{{\alpha }_{t,u}}V{_{t,u}}
\\&-{{\beta }_{t,u}}{{H}_{t,u}}-{{\gamma }_{t,u}}{R_{t,u}}],
\end{aligned}
\end{equation}
where  ${\mathbf{g}_{t,u}}=\left\{g_{k_{t,u}}^{l}\right\}_{l\in \mathcal{L}}$ 
and ${\mathbf{o}_{t,u}}=\left\{o_{k_{t,u}}^{l}\right\}_{l\in \mathcal{L}}$. Here,
${{\lambda }_{t,u}}$ is the relative memory length of user $u$ at scheduling slot~$t$, and ${p}_u$ is the requested video sequence length. Here, $\alpha_{t,u}$, $\beta_{t,u}$, and $\gamma_{t,u}$ are user $u$'s sensitivity degrees of QoE factors at scheduling slot $t$. A relatively small $\alpha_{t,u}$ indicates that user $u$ is not particularly concerned about video quality, while a large $\alpha_{t,u}$ means that more communication resources need to be allocated to transmit the segment of high video quality. A large $\beta_{t,u}$, relatively to the other parameters, indicates that user $u$ is deeply concerned about quality variation. The segment with suitable version needs to be transmitted to achieve smoother changes of video quality. In cases where user $u$ prefers low rebuffer time, a large $\gamma_{t,u}$ should be used.

\subsection{{DT-Assisted PQoE Model}}
To obtain user $u$'s PQoE model parameters $\lambda _{t,u}$, $\alpha_{t,u}$, $\beta_{t,u}$, $\gamma_{t,u}$ in real time, we utilize the UDT to analyze the user's data related to the adaptive video streaming. The user's engagement time is employed by UDT to calculate the objective PQoE reference value \cite{Dobrian}. Specifically, assume that a user only stops watching the video due to the long rebuffer time, low video quality, and frequent quality variation. If the proportion of the user's engagement time to the total video playback time is high, the user is insensitive to PQoE factors. Correspondingly, the PQoE is high; Otherwise, the PQoE is low. Therefore, the objective PQoE reference value is defined by
\begin{equation}
Z_{\hat{u},f}^{ref}=\frac{5\cdot q_{\hat{u},f}}{{{K}_{f}}\cdot e+{R_{\hat{u},f}}},
\end{equation}
where ${q_{\hat{u},f}}$ is the engagement time of user $\hat{u}$ watching video~$f$. Here, $K_f$ is the total number of segments of video~$f$, and~$R_{\hat{u},f}$ is the total rebuffer time for user $\hat{u}$ watching video $f$, which can be calculated based on the historical statistics on~$R_{t,\hat{u}}$. The PQoE range is from 0 to 5, aligning with the common principle \cite{Qiao_J}. To build the mapping relationship between~${Z}_{t,\hat{u}}$ and $Z_{\hat{u},f}^{ref}$, we need to accumulate the ${{Z}_{t,\hat{u}}}$ of video~$f$ in the time domain. Since the established PQoE model is a non-linear function, we employ the nonlinear regression method to obtain the PQoE model parameters. The input is the user's historical rebuffer time, video quality, quality variation, and objective PQoE reference value, and the output is PQoE model parameters in UDT, i.e., $\widetilde{\lambda}_{t,\hat{u}}$, $\widetilde{\alpha}_{t,\hat{u}}$, $\widetilde{\beta}_{t,\hat{u}}$, $\widetilde{\gamma}_{t,\hat{u}}$. Here, $\hat{u}$ corresponds to $u$. By substituting these model parameters into Eq. \eqref{QoE_1}, we can obtain the DT-assisted PQoE model, $\widetilde{Z}_{t,{u}}({\mathbf{g}_{t,{u}}},{\mathbf{o}_{t,{u}}},{{\omega }_{t,{u}}},{{\xi }_{t,{u}}})$.

\section{DRL-Based Resource Management Algorithm}\label{Solution}
\subsection{Problem Formulation}
Our objective is to maximize the overall long-term PQoE over $T$ scheduling slots. Correspondingly, the optimization problem is formulated as
\begin{align}\label{QoE}
{\text{P}_{0}}: &\underset{\begin{smallmatrix} 
	\{{\mathbf{g}_{t,{u}}},{\mathbf{o}_{t,{u}}},\\{\mathbf{\omega }_{t,{u}}},{\mathbf{\xi }_{t,{u}}}\}
	\end{smallmatrix}}{\mathop{\max }}\,\underset{T\to \infty }{\mathop{\lim }}\,\frac{1}{T}\sum\limits_{t=1}^{T}{\sum\nolimits_{{u}\in{{\mathcal{U}}}} {{\widetilde{Z}}_{t,{u}}}}({\mathbf{g}_{t,{u}}},{\mathbf{o}_{t,{u}}},{{\omega }_{t,{u}}},{{\xi }_{t,{u}}}) \\
	\text{s.t.}&  \sum\nolimits_{l=1}^{L}{g_{k_{t,{u}}}^{l}}\le 1, g_{k_{t,{u}}}^{l} \in \{0, 1\}, \tag{11{a}} \label{1a} \\ 
	& \sum\nolimits_{l=1}^{L}{o_{k_{t,{u}}}^{l}}\le 1, o_{k_{t,{u}}}^{l} \in \{0, 1\}, \tag{11{b}} \label{2a} \\
	& \sum_{l=1}^{L}{o_{k_{t,{u}}}^{l}\varsigma_{k_{t,{u}}}^{l}}\le \underset{l}{\mathop{\max }}\,\{\chi _{k_{t,{u}}}^{l}\varsigma_{k_{t,{u}}}^{l}, \forall l \in \mathcal{L}\}, \tag{11{c}} \label{2c}\\ 
	& \sum\nolimits_{{u}\in {\mathcal{U}}}{{\omega_{t,{u}}}\le 1}, \omega_{t,{u}} \in [0, 1], \tag{11{d}} \label{2b} \\ 
	& \sum\nolimits_{{u}\in {\mathcal{U}}}{{\xi_{t,{u}}}\le 1}, \xi_{t,{u}} \in [0, 1]. \tag{11{e}} \label{2d}
\end{align}
Constraint (\ref{1a}) indicates that a user can only receive the segment with one version at each scheduling slot. Constraint~(\ref{2a}) guarantees that a segment can only be transcoded for one target version for a user at each scheduling slot. Constraint~(\ref{2c}) guarantees that a segment can only be transcoded from the high version to the low version in the edge server at each scheduling slot. Constraint~(\ref{2b}) guarantees that the total allocated computing resources cannot exceed the computing capacity of the edge server. Constraint~(\ref{2d}) guarantees that total allocated bandwidth cannot exceed overall bandwidth. 

\subsection{Proposed Algorithm}
The formulated problem is mixed-integer nonlinear programming, which is hard to be directly solved. Considering that the user's playback status satisfies Markov chain and the optimization objective is to maximize the long-term performance, the optimization problem can be modeled as a Markov decision process (MDP). To solve this MDP, we adopt the deep deterministic policy gradient (DDPG) algorithm. 
\subsubsection{MDP}
MDP is a discrete-time stochastic control process, which consists of four elements, i.e., state, action, state transition probability, and reward. At each step (scheduling slot) $t$, state $s_t$ can be transformed to $s_{t+1}$ by taking action $a_t$. Correspondingly, the reward is $r_{t}$. 

\textbf{State:}
The state includes users' rebuffer time, buffer occupancy, segment version, video quality, and segment bitrate: 
\begin{equation}\small
{{s}_{t}}=\left\{{\left\{R_{t,{u}} \right\}_{{u}\in {\mathcal{U}}}},{\left\{B_{t, {u}} \right\}_{{u}\in {\mathcal{U}}}},{\left\{l_{t, {u}} \right\}_{{u}\in {\mathcal{U}}}}, {\left\{V_{t, {u}} \right\}_{{u}\in {\mathcal{U}}}}, \left\{\varsigma_{t,{u}} \right\}_{{u} \in {\mathcal{U}}} \right\}.
\end{equation}

\textbf{Action:} The action includes all optimization variables in Eq.~\eqref{QoE}, which is defined by
\begin{equation}\small
{{a}_{t}}=\left\{ \left\{\mathbf{g}_{t,{u}}\right\}_{{u}\in {\mathcal{U}}},\left\{\mathbf{o}_{t,{u}} \right\}_{{u}\in {\mathcal{U}}},\left\{ \omega _{t,{u}} \right\}_{{u}\in {\mathcal{U}}},\left\{\xi _{t,{u}} \right\}_{{u}\in {\mathcal{U}}} \right\}.
\end{equation}

\textbf{State Transition Probability:}
At each step, the current segment version selection, and communication and computing resource allocation affect user's next state. Therefore, the state transition probability between $s_t$ and $s_{t+1}$ is given by
\begin{equation}\small
	\begin{split}
		&\Pr ({{s}_{t+1}}|{{s}_{t}},{{a}_{t}})\!=\!\!\prod\limits_{u\in \mathcal{U}}{\Pr (R_{t+1,{u}}|R_{t,{u}},{\mathbf{g}_{t,{u}}},{\mathbf{o}_{t,{u}}},{{\omega }_{t,{u}}},{{\xi }_{t,{u}}})} \cdot  \\ 
		&\prod\limits_{{u}\in {\mathcal{U}}}{\Pr (V_{t+1,{u}}|V_{t,{u}},{\mathbf{g}_{t,{u}}})}\cdot \prod\limits_{{u}\in {\mathcal{U}}}{\Pr (\varsigma_{t+1,{u}}|\varsigma_{t,{u}},{\mathbf{g}_{t,{u}}},{\mathbf{o}_{t,{u}}})} \cdot  \\ 
		&\prod\limits_{{u}\in {\mathcal{U}}}{\Pr (B_{t+1,{u}}|B_{t,{u}},{\mathbf{g}_{t,{u}}})} \cdot \prod\limits_{{u}\in {\mathcal{U}}}{\Pr (l_{t+1,{u}}|l_{t,{u}},{\mathbf{g}_{t,{u}}})}.
	\end{split}
\end{equation}

\textbf{Reward:} 
The reward at step $t$ is designed to maximize overall PQoE, which is defined by 
\begin{equation}\label{r}
r_t(s_t, a_t) = \sum\nolimits_{{u}\in{{\mathcal{U}}}}\widetilde{Z}_{t,{u}}.
\end{equation}

\subsubsection{\underline{D}RL-Based \underline{C}ommunication and \underline{T}ranscoding \underline{R}esource \underline{A}llocation (DCTRA) Algorithm}
Based on DT-assisted PQoE models and formulated MDP, we adopt the DDPG algorithm to obtain segment version selection, and communication and computing resource allocation decisions \cite{Tim}\cite{Zhou}. The training process is as shown in Algorithm \ref{alg1}. The DDPG contains three modules, i.e., primary network, target network, and replay memory. Primary network consists of one actor network, i.e., $\pi(s_t|\theta_{\pi})$, and one critic network, i.e., $Q(s_t, a_t|\theta_{Q})$, where $\theta_{\pi}$ and $\theta_{Q}$ are primary actor and critic network parameters, respectively. Primary network is mainly responsible for finding the best action $a_t$ for the current state $s_t$. The target network is the same as the primary network in terms of network structure but with different parameters $\theta_{\pi'}$ and $\theta_{Q'}$, which is responsible for generating target values for training the primary actor and critic networks. At each step $t$, $(s_t, a_t, s_{t+1}, r_t)$ is stored at replay memory $M$ as an experience tuple for random sampling. The size of random sampling is the mini-batch. In each step, the ornstein-uhlenbeck noise added to action is denoted by $X_t$. The network learning rate is denoted by $\tau$.

The primary actor network parameter, i.e., $\theta_{\pi}$, is updated based on the sampled policy gradient, as follows
\begin{equation}\label{gradient}
\bigtriangledown_{\theta_{\pi}}J\approx\frac{1}{N}\sum_t\bigtriangledown_a Q(s,a|\theta_Q)|_{s=s_t,a=\pi(s_t)} \bigtriangledown_{\theta_{\pi} \pi(s|\theta_{\pi})|_{s=s_t}
} ,
\end{equation}
where $Q(s,a|\theta_Q)$ is the action-value function. The parameter $\theta_Q$ of primary critic network is updated by minimizing the loss function ${G}(\theta_Q)$ as follows
\begin{equation}\label{loss}
{G}(\theta_Q) = \frac{1}{N}\sum_t(y_t - Q(s_t,a_t|\theta_Q))^2.
\end{equation}
Here, $y_t$ is the target value that combines the current reward and the estimated Q value, i.e., $Q'$, as follows
\begin{equation}\label{y_t}
	y_t = r_t + \epsilon Q'(s_{t+1},\pi'(s_{t+1}|\theta_{\pi'})|\theta_{Q'}),
\end{equation}
where $\epsilon$ is the discounting factor.

Since the output values of the DDPG algorithm are continuous, we need to discretize a part of DDPG output values to obtain $\mathbf{g}_{t,{u}}$ and $\mathbf{o}_{t,{u}}$\label{6}. To reduce the dimension of $\mathbf{g}_{t,{u}}$ and $\mathbf{o}_{t,{u}}$ caused by multiple segment versions, we utilize $g_{t,u}$ and $o_{t,u}$ to replace $\mathbf{g}_{t,{u}}$ and $\mathbf{o}_{t,{u}}$. Specifically, we first map the output value of the Tanh function from $[-1, 1]$ to $[0, 1]$. Then, we split the range of ${g}_{t,{u}}$ into five parts, i.e., $[0, 0.2), [0.2, 0.4), [0.4, 0.6), [0.6, 0.8), [0.8, 1]$, which indicate no segment transmission, and segment transmission for version 1, 2, 3, 4 for user ${u}$ at step $t$, respectively. Similarly, the range of ${o}_{t,{u}}$ is split into two parts, i.e., $[0, 0.5), [0.5, 1]$, which indicate no segment transcoding, and segment transcoding for user ${u}$ at step $t$, respectively.

\begin{algorithm}[t]
	\caption{DCTRA}
	\label{alg1}
	\textbf{Initialize:} primary and target network with parameters $\theta_{\pi}, \theta_{Q}, \theta_{\pi'}, \theta_{Q'}$, and replay memory $M$.
	
	\For{each episode}
	{
		Initialize video list in edge server and cloud server, users’ request videos, buffer queues and locations.
		
		\For{each step $t \in \{1, ..., t_{max}\}$}
		{
			Update each PQoE model based on parameters $\widetilde{\lambda}_{t,{u}}$, $\widetilde{\alpha}_{t,{u}}$, $\widetilde{\beta}_{t,{u}}$, $\widetilde{\gamma}_{t,{u}}$ provided by UDTs;
			
			Discretize actions $g_{t,u}$ and $o_{t,u}$ based on the designed principle in Section  \ref{6};
			
			BS and edge server execute action $a_t$ based on $a_t = \pi(s_t|\theta_{\pi}) + {X}_t$;

			Obtain $r_t$ based on Eq. \eqref{r}, update $s_{t+1}$;
							
			Store $(s_t, a_t, r_t, s_{t+1})$ into $M$, and sampling;
			
			Compute tagret value $y_t$ based on Eq. \eqref{y_t};
			
			Update primary critic network parameter $\theta_Q$ by minimizing Eq. \eqref{loss};
			
			Update primary actor network parameter $\theta_{\pi}$ by minimizing Eq. \eqref{gradient};
			
			Update target network parameters via 
			$\theta_{\pi'} \leftarrow \tau\theta_{\pi} + (1-\tau)\theta_{\pi'},\theta_{Q'} \leftarrow \tau\theta_{Q} + (1-\tau)\theta_{Q'}$;
		}
	}
\end{algorithm}

\section{Performance Evaluation}\label{Result}

In this section, simulations are conducted to evaluate the performance of the proposed DCTRA algorithm. We adopt the real-world dataset ``Waterloo Streaming Quality-of-Experience Database-III"\footnote{https://ieee-dataport.org/open-access/waterloo-streaming-quality-experience-database-iii.}, which includes users' rebuffer time, video quality and quality variation. The dataset contains 450 videos, and the length of each video is 10 seconds. The user's leave probability in \cite{Nam} is used to obtain the engagement time for calculating the PQoE reference value. To obtain PQoE model parameters, we employ the lsqcurvefit function\footnote{https://www.mathworks.com/help/optim/ug/lsqcurvefit.html.} in Matlab to fit the function curve. The main simulation parameters are presented in Table \ref{Table1}. 

\begin{table}[t]
	\footnotesize
	\centering
	\captionsetup{justification=centering, singlelinecheck=false}
	\caption{Simulation Parameters}
	\label{Table1}
	\begin{tabular}{c c|c c}
		\hline
		\hline
		\textbf{Parameter}                 & \textbf{Value} & \textbf{Parameter}        & \textbf{Value}  \\ \hline
		BS radius                          & 600 m               &    $U$            & 12 \\ 
		Hidden layer shape & (512, 256, 128)            & $c$  &   $10^9$ cycles/s          \\ 
		Critic learning rate                          &      $10^{-4}$      &  $W$   & 200 MHz          \\ 
		Actor learning rate       & $10^{-6}$ &	$r^{C}$                            & 200 Mbps                   \\ 
		Mini-batch size               & 128 & $P$                          & 25 dBm                     \\ 
		Replay memory & $6\times10^3$ & $\mu$  & 10 cycles/bit   \\
		Scheduling slot length & 100 ms & $L$ & 4 \\
		Number of episodes  & 1500  &  $t_{\text{max}}$  &  100 \\
		 \hline
	\end{tabular}
\end{table}

We compare the proposed DTCRA algorithm with the following benchmark schemes. 
\begin{itemize}
	\item[$\bullet$] \textbf{\underline{R}ound-\underline{R}obin (RR)}: In each scheduling slot, communication and computing resources are equally allocated to 3 users randomly selected from 12 users. The segment version is randomly selected from all versions.
	
	\item[$\bullet$] \textbf{\underline{P}roportional \underline{F}air (PF)}: Resources are allocated based on the scheduling priority considering user's current channel state and buffer occupancy. The segment version is selected based on the previous average version.
	
	\item[$\bullet$] \textbf{\underline{J}oint \underline{R}esource \underline{A}lloca\underline{T}ion (JRAT)}\cite{Huang_X_1}: 
	The communication and transcoding resources are sequentially allocated to the user who can achieve the highest PQoE enhancement. The segment version selection is determined by the branch and bound method.
	
	\item[$\bullet$] \textbf{\underline{C}ommunication and \underline{T}ranscoding \underline{R}esource \underline{A}llocation (CTRA):} The proposed DCTRA algorithm does not use DT-assisted PQoE models, but uses static PQoE models.
\end{itemize}

\begin{figure}[t]
	\centering
	\includegraphics[width=8.5cm]{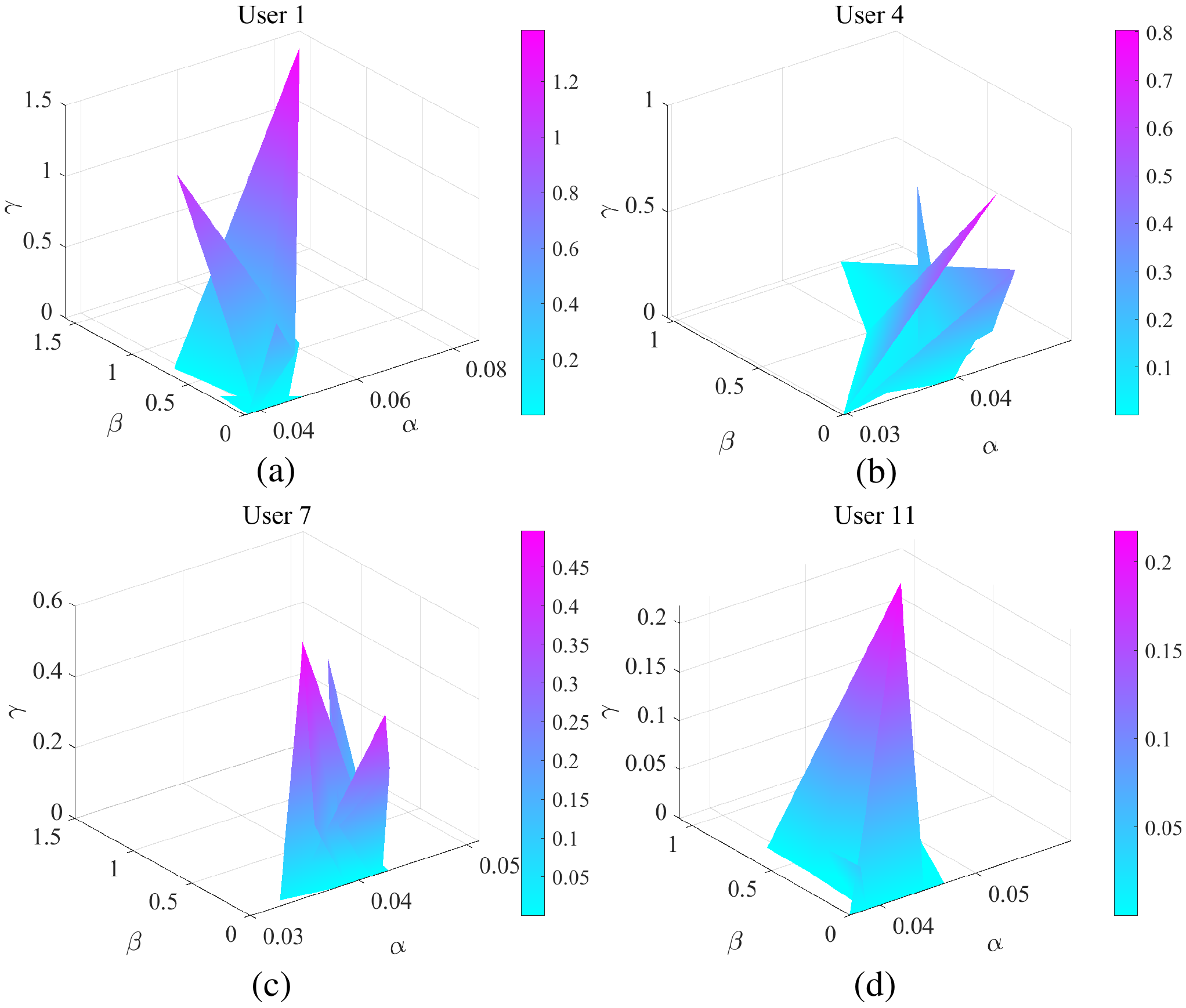}
	\caption{PQoE model parameters extracted by UDT.}
	\label{fig:Feature}
\end{figure}

Figure \ref{fig:Feature} shows the dynamic PQoE model parameters of Users 1, 4, 7, and 11, whose average relative memory lengths are $12.80$ seconds, $11.93$ seconds, $2.02$ seconds, $11.68$ seconds, respectively. It can be observed that User $1$ has the highest sensitivity degree of all QoE factors, while User $11$ owns the lowest one. In addition, the user's sensitivity degree of QoE factors changes with varying video contents and playback statuses. 

\begin{figure*}\label{fig4}
	\centering
	\subfigure[]{
		\label{fig4:first}
		\includegraphics[width=0.32\textwidth]{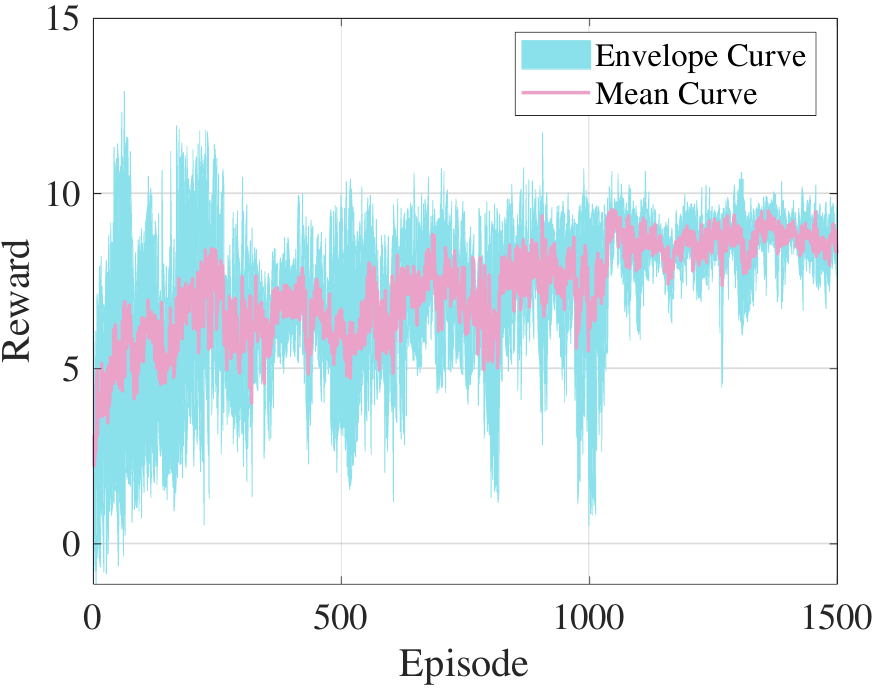}}
	\centering
	\subfigure[]{
		\label{fig4:second}
		\includegraphics[width=0.32\textwidth]{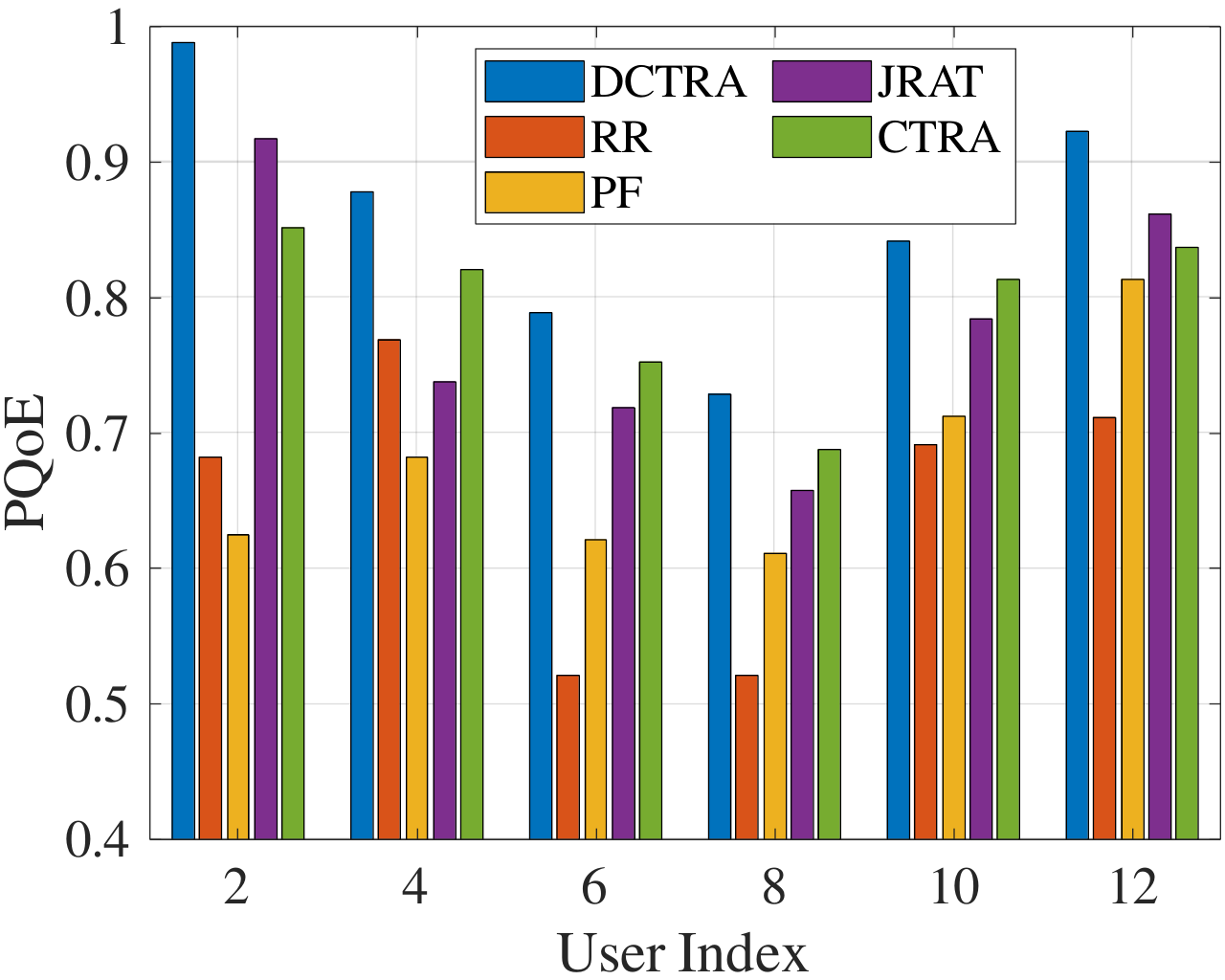}}
	\centering
	\subfigure[]{
		\label{fig4:third}
		\includegraphics[width=0.32\textwidth]{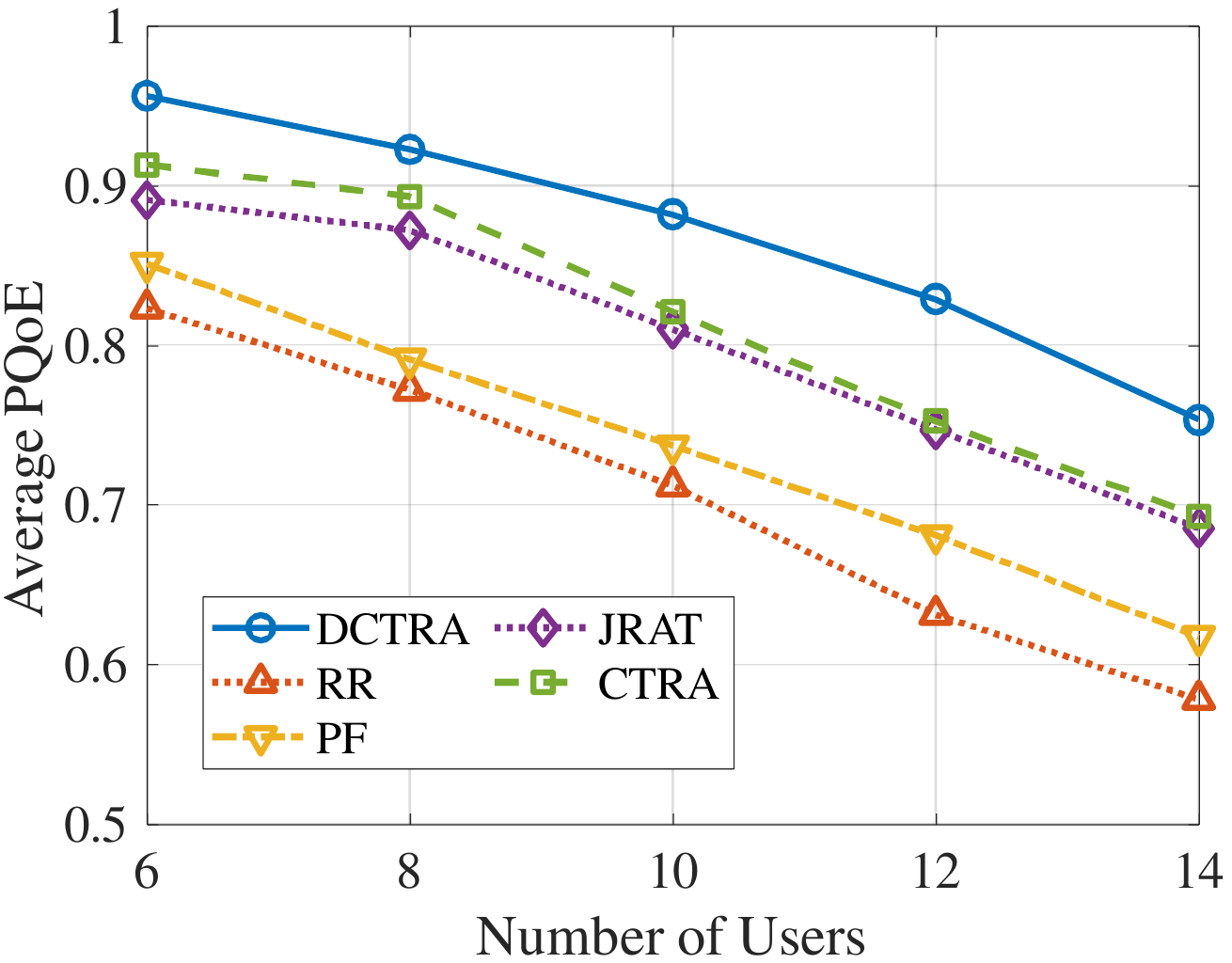}}
		\caption{Performance comparison among different algorithms: (a) convergence performance of the DCTRA algorithm, (b) PQoE (normalized) of different users, and (c) average PQoE (normalized)  among different numbers of users.}
	\label{fig4}
\end{figure*}

Figure \ref{fig4:first} shows the convergence property of the proposed DCTRA algorithm, which conducts three trials of training to draw the corresponding envelope curve and the mean curve. It can be observed that our algorithm can converge after near 1,100 episodes. In Fig. \ref{fig4:second}, we select 6 users from 12 users to show the individual user's normalized PQoE among different algorithms. It can be observed that the proposed DCTRA algorithm can always achieve the highest PQoE for each user compared with other algorithms. This is because accurate user-specific PQoE models can be provided to the network controller in real time to perform appropriate resource management. Fig.~\ref{fig4:third} shows the proposed algorithm can well adapt to different user scales and always achieve the highest PQoE. The gap between the JRAT algorithm and the CTRA algorithm decreases gradually, because these two algorithms cannot find better resource management decisions based on static PQoE models.

\section{Conclusion}\label{Conclusion}
In this paper, we have investigated a PQoE enhancement problem for adaptive video streaming. We have proposed a DT-assisted PQoE model construction method to obtain accurate user-specific PQoE models, and then developed a DRL-based resource management algorithm to enhance the overall long-term PQoE. The proposed scheme can extract users' features through DT, and leverage them to enhance users' video watching experience. In the future, we will investigate an efficient data collection and model update scheme for UDT to reduce communication and computing overhead.



\bibliographystyle{IEEEtran}
\bibliography{Ref}

\end{document}